\documentclass[aps,prl,twocolumn,footinbib,10pt]{revtex4-1} 

\usepackage{caption}
\usepackage{graphicx}
\usepackage{float}
\usepackage{braket}
\usepackage{color}
\usepackage{ragged2e}
\usepackage{mathtools}
\usepackage{amsmath}
\usepackage{amssymb}
\usepackage{hyperref}
\usepackage{float}
\usepackage{epstopdf}

\begin{document}

\title{Screening of the quantum dot F\"orster coupling at small distances}

\author{Chelsea Carlson$^{1*}$}
\author{Andreas Knorr$^{2}$}
\author{Stephen Hughes$^{1}$}

\affiliation{$^{1}$Department of Physics, Queen's University, Kingston, Ontario, Canada, K7L 3N6}

\affiliation{$^{2}$Institute of Theoretical Physics, Nonlinear Optics and Quantum Electronics, Technical University of Berlin, 10623 Berlin, Germany}

\affiliation{$^{*}$Corresponding author: 0clac@queensu.ca}

\date{\today}

\begin{abstract} 

We study the near-field energy transfer rates between two finite size quantum dot disks, generalizing the result of F\"orster coupling between two point dipoles. In particular, we derive analytical results for the envelope of the electronic wavefunction for model potentials at the boundaries of  quantum dot disks and demonstrate how the F\"orster interaction is screened as the size of the dots becomes comparable to the dot-dot separation. 

\end{abstract}

\maketitle

Semiconductor quantum emitters play a key role in many applications in optoelectronics, quantum electodynamics (QED), and fundamental optical science. Although atomic or molecular dipole emitters are commonly studied for their clear two- or multi-level energy transitions which are easy to understand theoretically, they are not suitable for integration with mature semiconductor nanotechnologies
not for frequency tuning of the resonant transition energies.
In contrast,
semiconductor quantum dots (QDs) are highly tunable and controllable 
``artificial atoms.'' Unlike true atoms, which fall well within the point dipole approximation (PDA) in the presence of visible light, QDs can have sizes, and distances with respect to each other, that can vary from several nanometers to tens or hundereds of nanometers \cite{dalacu_selective_2011}, which is not only non-negligible compared to the typical wavelengths of light but is highly important when integrated in nanophotonic structures that mould the flow and confinement of light on similar size scales. 
Even for single quantum emitters,  for QD emitters and large molecules having comparable size and distance, both experimental and theoretical analysis have shown that the PDA is no longer valid due to the finite size of the dipole \cite{ahn_radiative_2003, stobbe_spontaneous_2012, cotrufo_spontaneous_2015, neuman_coupling_2018}. 
Further examples where PDA breaks down include QD emitters coupled to a nanoplasmonic heterostructure, where the photonic local density of states (LDOS) can vary wildly on the single- and sub-nanometer scale \cite{vanvlack_finite-difference_2012, chikkaraddy_single-molecule_2016}. 



Significant work has been done to grow and characterize coupled pairs of vertically stacked QDs \cite{bayer_coupling_2001, gerardot_photon_2005, dalacu_selective_2011, bartolome_strain_2017}, as well as supporting theory investigations \cite{danckwerts_theory_2006, carlson_theory_2019}, where one of the prominent features of the coupled-QD spectra is a large energy splitting ($>$meV) of the exciton transition energy, commonly known and treated as F\"{o}rster coupling \cite{forster_zwischenmolekulare_1948} (if outside the regime of electronic tunnelling). 
To move beyond the PDA, several works have implemented a wavefunction approach to describe the QD system at the single-QD level \cite{stobbe_spontaneous_2012} as well as multi-QD level \cite{rozbicki_quantum_2007}. For cylindrical QDs, like those in a stacked configuration, the solution to the simple harmonic oscillator (SHO) potential is typically used to approximate the electron-hole (exciton) wavefunction in-plane, and either the SHO or the infinite square well (ISW) potential along the QD axis. Other geometries of QDs, like spherical, have been previously described using superpositions of atomic potentials \cite{scholes_resonance_2005}. 

Partly motivated by emerging experiments~\cite{carlson_theory_2019},
in this work, we generalize the PDA between two quantum dots (the usual F\"{o}rster coupling between point dipoles) to systems where the distance between two dots can be as small as their extension. We use a two-space Coloumb Green function as well as the electronic wavefunction for a thin-disk quantum dot geometry. The wavefunction is evaluated using several different electronic potentials including the SHO and the infinite square well, as well as approximating the wavefunction as a step function over the area of the QD. We use these results to calculate the effective F\"{o}rster potential between two vertically stacked QDs separated by a small gap, and thus, the expected effective splitting of the electronic energy level. More notably, we show that by approximating the wavefunction as a simple step function and assuming infinitesimally thin QDs, we can simplify the 
difficult 6-dimensional spatial integral with a very simple 1-dimensional integral, and achieve a reasonable approximation to the more complex example of a finite-thickness QD using a SHO potential. In the limit of large QD gaps,  we also show how and when the PDA is recovered.

\begin{figure}[h!]
    \centering
    \includegraphics[width=\columnwidth]{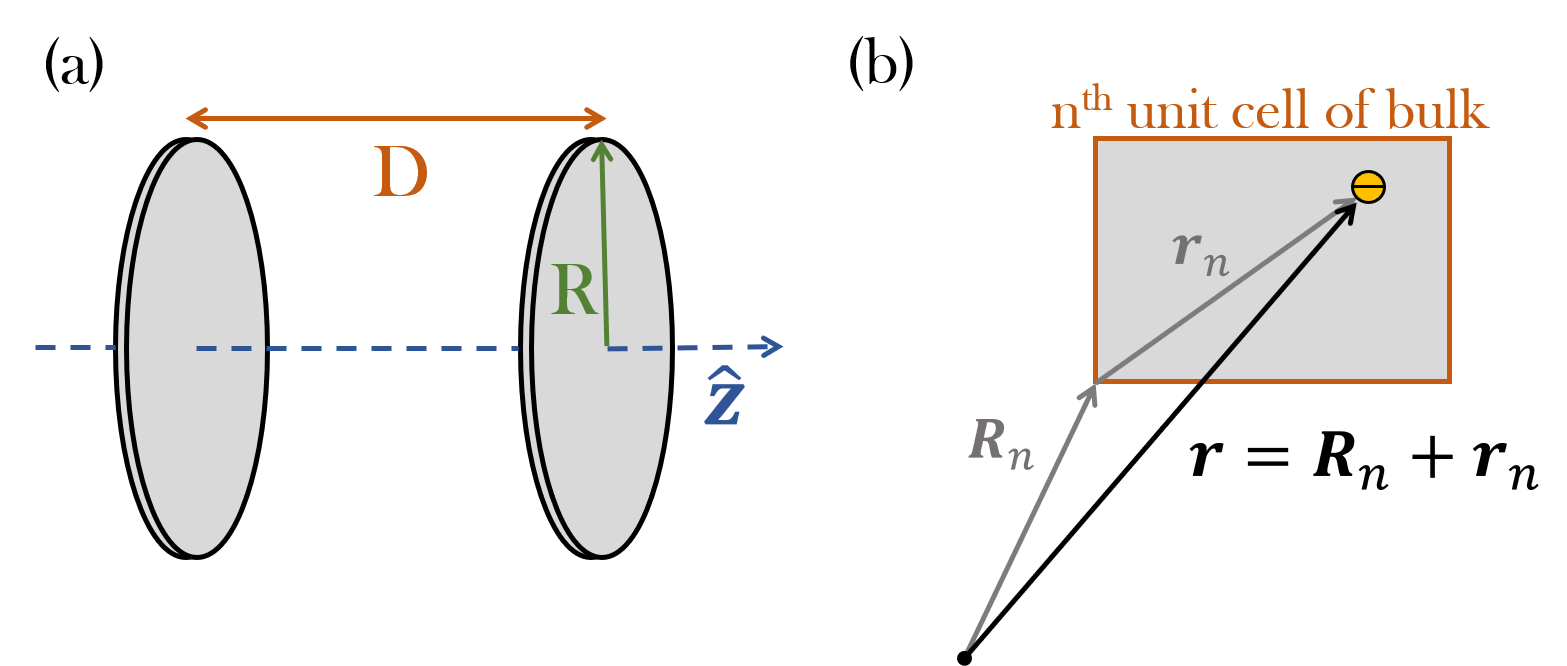}
    \caption{(a) Schematic of coupled QDs and (b) coordinate system. }
    \label{fig:schematic_dots}
\end{figure}

Using a longitudinal near-field interaction, we can write the interaction Hamiltonian for the two QD dots 
as~\cite{richter_theory_2006,zimmerman_poisson_2016}:
\begin{equation}
    \mathcal{H} = \frac{1}{2}\sum_i \phi(\mathbf{r}_i)q_i = \frac{1}{2}\int{\rm d}\mathbf{r} \phi(\mathbf{r})\rho(\mathbf{r}),
    \label{eq:Hamiltonian}
\end{equation}
where $q$ is the carrier charge, $\rho$ is the charge density, and $\phi$ is the electromagnetic potential. Figure~\ref{fig:schematic_dots} shows the setup of the two dots, where $D$ is the centre-to-centre distance between the dots and $R$ is the radius of the dots, as well as the coordinate system used in the rest of the formulation; here, $\mathbf{r}_n$ is the coordinate of the charge carrier within the unit cell for the bulk material of the quantum dots (i.e., InAs), and $\mathbf{R}_n$ is the coordinate of the unit cell, resulting in the total vector $\mathbf{r} = \mathbf{R}_n + \mathbf{r}_n$. 
For the wavefunction of electrons, we will use an envelope approximation such that,    $\psi^a_\lambda(\mathbf{r}) = \zeta^a_\lambda(\mathbf{R}_n) u_\lambda(\mathbf{r}_n)$, where $a$ is the dot number (1 or 2), $\lambda$ is the band number (valence or conduction), $\zeta^a_\lambda(\mathbf{R}_n)$ is the envelope function, and $u_\lambda(\mathbf{r}_n)$ is the Bloch function within the $\rm n^{th}$ unit cell.  The potential is given by
\begin{equation}
    \mathbf{\nabla}\cdot\varepsilon(\mathbf{r})\mathbf{\nabla}\phi(\mathbf{r}) = -\frac{\rho(\mathbf{r})}{\varepsilon_0},
    \label{eq:phi}
\end{equation}
where $\varepsilon$ is the relative electric-permittivity (dielectric constant) of the bulk material of the QDs and $\varepsilon_0$ is the vacuum permittivity. Later on, we simplify the system to a homogeneous medium in the near field, $\varepsilon(\mathbf{r}) \rightarrow \varepsilon~(\simeq$12 for InAs), and Eq.~\ref{eq:phi} can be simplified to $\mathbf{\nabla}^2\phi(\mathbf{r}) = -\frac{\rho(\mathbf{r})}{\varepsilon\varepsilon_0}$. More generally, We can write the Coloumb Green function in a similar fashion, 
defined from
\begin{equation}
    \mathbf{\nabla}\cdot\varepsilon(\mathbf{r})\mathbf{\nabla}{G}(\mathbf{r},\mathbf{r}^\prime) = -\delta(\mathbf{r}-\mathbf{r}^\prime)\frac{1}{\varepsilon_0}.
    \label{eq:GF}
\end{equation}
The Green function is easily obtainable either analytically for a homogeneous medium (or simple geometries like waveguides) or using numerical methods for more complicated geometries. In our coordinate system and within the rotating wave approximation, we may approximate the Green function into two parts, namely monopole-monopole and dipole-dipole interactions, using a Taylor series expansion, since $\mathbf{R}_n>>\mathbf{r}_n$:
\begin{equation}
\begin{aligned}
    {G}(\mathbf{r},\mathbf{r}^\prime) =& 
    {G}(\mathbf{R}_n+\mathbf{r}_n,\mathbf{R}_{n^\prime}+\mathbf{r}_{n^\prime}) \\
    \simeq& {G}(\mathbf{R}_n,\mathbf{R}_{n^\prime}) +
    (\mathbf{r}_n\cdot\mathbf{\nabla}_{\mathbf{R}_n})
    (\mathbf{r}_{n^\prime}\cdot\mathbf{\nabla}_{\mathbf{R}_{n^\prime}}){G}(\mathbf{R}_n,\mathbf{R}_{n^\prime}) + ... ,
\end{aligned}
\label{eq:GF-expanded}
\end{equation}
where the first term includes monopole interactions, the second term included dipole interactions, and other terms include higher order interactions.  The interaction Hamiltonian (Eq.~\ref{eq:Hamiltonian}) can now be written in terms of the envelope function, Bloch function, and Green function using $\phi(\mathbf{r}) = \int{\rm d}\mathbf{r}^\prime G(\mathbf{r},\mathbf{r}^\prime)\rho(\mathbf{r}^\prime)$ as well as the second-quantization of the charge density as an expansion of electronic eigenstates, $\rho(\mathbf{r}) = q\sum_{aa^\prime}a^\dagger_a a_{a^\prime}\psi^*_a(\mathbf{r})\psi_{a^\prime}(\mathbf{r})$:
\begin{equation}
\begin{aligned}
\mathcal{H} =& \frac{1}{2}\int{\rm d}\mathbf{r}\phi(\mathbf{r})\rho(\mathbf{r})=  \frac{q^2}{2}\sum_{a,a^\prime,b,b^\prime}a^\dagger_a a_{a^\prime}a^\dagger_b a_{b^\prime} \times
\\
&\int{\rm d}\mathbf{r}\int{\rm d}\mathbf{r}^\prime
\psi^*_a(\mathbf{r})\psi_{a^\prime}(\mathbf{r})
G(\mathbf{r},\mathbf{r}^\prime)
\psi^*_b(\mathbf{r}^\prime)\psi_{b^\prime}(\mathbf{r}^\prime),\\
=& \frac{1}{2}\sum_{a,a^\prime,b,b^\prime}a^\dagger_a a_{a^\prime}a^\dagger_b a_{b^\prime}
    V^{aa^\prime bb^\prime}_F,
\end{aligned}
\label{eq:Hamiltonian-expanded}
\end{equation}
where the indices on the creation/annihilation operators ($a_i^\dagger/a_i$) represent all quantum numbers present such as dot number and band number, and $V_F^{aa^\prime bb^\prime}$ is the effective potential. 

The effective potential can then be written (using only the dipole-dipole interaction term in 
Eq.~\ref{eq:GF-expanded}), 
\begin{equation}
\begin{aligned}
V^{aa^\prime bb^\prime}_F 
=& 
q^2\int{\rm d}\mathbf{r}\int{\rm d}\mathbf{r}^\prime
\psi^*_a(\mathbf{r})\psi_{a^\prime}(\mathbf{r})
G(\mathbf{r},\mathbf{r}^\prime)
\psi^*_b(\mathbf{r}^\prime)\psi_{b^\prime}(\mathbf{r}^\prime),\\
\simeq& 
q^2\sum_{n,n^\prime}\int{\rm d}\mathbf{r}_n\int{\rm d}\mathbf{r}_{n^\prime}
\zeta^*_a(\mathbf{R}_n)\zeta_{a^\prime}(\mathbf{R}_n)u^*_a(\mathbf{r}_n)u_{a^\prime}(\mathbf{r}_n)\\
&\hspace{2cm}[(\mathbf{r}_n\cdot\mathbf{\nabla}_{\mathbf{R}_n})
(\mathbf{r}_{n^\prime}\cdot\mathbf{\nabla}_{\mathbf{R}_{n^\prime}}){G}(\mathbf{R}_n,\mathbf{R}_{n^\prime})]\\
&\hspace{2cm}\zeta^*_b(\mathbf{R}_{n^\prime})\zeta_{b^\prime}(\mathbf{R}_{n^\prime})u^*_b(\mathbf{r}_{n^\prime})u_{b^\prime}(\mathbf{r}_{n^\prime}),\\
=& 
q^2\sum_{n,n^\prime}
\int{\rm d}\mathbf{r}_nu^*_a(\mathbf{r}_n)u_{a^\prime}(\mathbf{r}_n)(\mathbf{r}_n\cdot\mathbf{\nabla}_{\mathbf{R}_n})\\
&\hspace{0.9cm}\int{\rm d}\mathbf{r}_{n^\prime}u^*_b(\mathbf{r}_{n^\prime})u_{b^\prime}(\mathbf{r}_{n^\prime})(\mathbf{r}_{n^\prime}\cdot\mathbf{\nabla}_{\mathbf{R}_{n^\prime}})
{G}(\mathbf{R}_n,\mathbf{R}_{n^\prime})\\
&\hspace{1.5cm}
\zeta^*_a(\mathbf{R}_n)\zeta_{a^\prime}(\mathbf{R}_n)\zeta^*_b(\mathbf{R}_{n^\prime})\zeta_{b^\prime}(\mathbf{R}_{n^\prime}),
\end{aligned}
\label{eq:potential-expanded}
\end{equation}
which gives a generalized F\"{o}rster coupling for the dipole-dipole interaction including the limit of the usual F\"{o}rster coupling. The interband dipole moment on the microscopic scale of the bulk material can be defined as:
\begin{equation}
\mathbf{d}_{ab} = \frac{1}{V_{UC}}\int{\rm d}\mathbf{r}_n u^*_a(\mathbf{r}_n)u_{b}(\mathbf{r}_n)\mathbf{r}_n,    
\label{eq:dipole}
\end{equation}
where ${V_{UC}}$ is the volume of the unit cell. Utilizing Eq.~\ref{eq:dipole} and changing from discrete variables to continuous ($\sum_n \rightarrow \frac{1}{V_{UC}}\int{\rm d}\mathbf{r}$), Eq.~\ref{eq:potential-expanded} becomes
\begin{equation}
\begin{aligned}
V^{aa^\prime bb^\prime}_F 
=& 
\int {\rm d}\mathbf{r}\int {\rm d}\mathbf{r}^\prime
\bigg[(\mathbf{d}_{aa^\prime}\cdot\mathbf{\nabla}_{\mathbf{r}})
(\mathbf{d}_{bb^\prime}\cdot\mathbf{\nabla}_{\mathbf{r}^\prime})
{G}(\mathbf{r},\mathbf{r}^\prime)\bigg]\\
&\hspace{1.8cm}\zeta^*_a(\mathbf{r})\zeta_{a^\prime}(\mathbf{r})\zeta^*_b(\mathbf{r}^\prime)\zeta_{b^\prime}(\mathbf{r}^\prime),\\
=& 
\int {\rm d}\mathbf{r}\int {\rm d}\mathbf{r}^\prime
\bigg[
\mathbf{d}_{aa^\prime}\cdot\mathbf{\nabla}_{\mathbf{r}}\bigg(\zeta^*_a(\mathbf{r})\zeta_{a^\prime}(\mathbf{r})\bigg)\\
&\hspace{1.8cm}\mathbf{d}_{bb^\prime}\cdot\mathbf{\nabla}_{\mathbf{r}^\prime}\bigg(\zeta^*_b(\mathbf{r}^\prime)\zeta_{b^\prime}(\mathbf{r}^\prime)\bigg)
\bigg]
{G}(\mathbf{r},\mathbf{r}^\prime),
\end{aligned}
\label{eq:potential-expanded2}
\end{equation}
which is a 6-d integral (4-d if we consider the dots to be infinitely thin) that can be numerically quite challenging. 
%
%
%

{\it Analytical Results.---}Before applying numerical calculations, we aim to simplify both the geometry and mathematics of the solution presented in Eq.~\ref{eq:potential-expanded2} by taking the thickness of the dots to be infinitesimally thin:
\begin{equation}
\begin{aligned}
   &{\rm Dot~1:~} \zeta^*_v(\mathbf{r})\zeta_c(\mathbf{r}) = \delta(z)\frac{\Theta(R - \rho)}{\pi R^2},\\
   &{\rm Dot~2:~} \zeta^*_v(\mathbf{r}^\prime)\zeta_c(\mathbf{r}^\prime) = \delta(z^\prime-D)\frac{\Theta(R - \rho^\prime)}{\pi R^2},
\end{aligned}
\end{equation}
where the dimensions are shown in Fig.~\ref{fig:schematic_dots} and $\rho$ is the radial direction in cylindrical coordinates (and angular direction will be denoted by $\varphi$). The second approximation is that the Green function is dominantly given by the homogeneous solution in the near field, even though waveguides can add a significant contribution\cite{carlson_theory_2019}. 
The homogeneous Green function can be derived from Eq.~\ref{eq:GF} by considering the permittivity to be spatially independent, $\varepsilon(\mathbf{r}) = \varepsilon$, and is given by
\begin{equation}
\begin{aligned}
   G(\mathbf{r},\mathbf{r}^\prime) = \frac{1}{4\pi\varepsilon_0\varepsilon\vert \mathbf{r}-\mathbf{r}^\prime\vert}.
\end{aligned}
\label{eq:gf-homo}
\end{equation}
Starting from Eq.~\ref{eq:potential-expanded2}, and assuming the dipoles are oriented in the $x$ direction (or $y$), and using the homogeneous Green function given by Eq.~\ref{eq:gf-homo},
\begin{equation}
\begin{aligned}
V^{aa^\prime bb^\prime}_F 
\!=& 
\frac{d^2}{4\pi\varepsilon_0\varepsilon \varepsilon}\int {\rm d}\mathbf{r}\int {\rm d}\mathbf{r}^\prime\\
&~~~~\times\frac{
\bigg(\nabla_x\zeta^*_v(\mathbf{r})\zeta_c(\mathbf{r})\bigg)
\bigg(\nabla_{x^\prime}\zeta^*_v(\mathbf{r}^\prime)\zeta_c(\mathbf{r}^\prime)\bigg)}
{\vert \mathbf{r}-\mathbf{r}^\prime\vert}.
\end{aligned}
\label{eq:potential-matlab-1}
\end{equation}

Converting to cylindrical coordinates, such that $(\mathbf{d}\cdot\mathbf{\nabla}_{\mathbf{r}})\Theta(R-\rho) = -d\delta(R-\rho)\cos{(\varphi)}$, 
%
%
the effective potential can be re-written as (for $\mathbf{d}_{aa^\prime} = d\hat{\mathbf{x}}$ and $\mathbf{d}_{bb^\prime} = d\hat{\mathbf{x}}^\prime$ ):
\begin{equation}
\begin{aligned}
V^{aa^\prime bb^\prime}_F 
=&
\frac{d^2}{4\pi\varepsilon_0\varepsilon}\frac{1}{(\pi R^2)^2}
\int_{\rm 6D}{\rm d}\rho{\rm d}\varphi{\rm d}z{\rm d}\rho^\prime{\rm d}\varphi^\prime{\rm d}z^\prime\\
&\frac{\rho\rho^\prime\delta(z)\delta(z^\prime-D)\nabla_{x}\nabla_{x^\prime}\big(\Theta(R - \rho)\Theta(R - \rho^\prime)\big)}
{[\rho^2+\rho^{\prime 2}-2\rho\rho^\prime\cos{(\varphi-\varphi^\prime)}+({z-z^\prime})^2]^{1/2}}
,\\
=&
\frac{d^2}{4\pi^3\varepsilon_0\varepsilon R^2D}
\iint {\rm d}\varphi{\rm d}\varphi^\prime \frac{\cos{(\varphi)}\cos{(\varphi^\prime)}}{[4\alpha^2\sin^2{(\frac{\varphi-\varphi^\prime}{2})}+1]^{1/2}},
\end{aligned}
\label{eq:potential-simplifiedfunction}
\end{equation}
where we have defined:

\begin{equation}
    \alpha\equiv R/D.
\end{equation} 

To separate $\varphi$ and $\varphi^\prime$ in the integrand, let us choose a new coordinate system for which we define $\varphi_- = \varphi-\varphi^\prime$ and $\varphi_+ = \varphi+\varphi^\prime$. In this new coordinate system, 
%
$\cos{(\varphi)}\cos{(\varphi^\prime)} = \frac{1}{2}\big(\cos{(\varphi_-)} + \cos{(\varphi_+)}\big)$,
%
and,
%
$\sin^2{\big(\frac{\varphi-\varphi^\prime}{2}\big)}= \sin^2{\big(\frac{\varphi_-}{2}\big)}$.
%
Thus, Eq.~\ref{eq:potential-simplifiedfunction} becomes,
\begin{equation}
\begin{aligned}
V_F=&
\frac{d^2}{4\pi^3\varepsilon_0\varepsilon R^2D}
\int {\rm d}\varphi_-\int {\rm d}\varphi_+ \frac{\frac{1}{2}\big[\cos{(\varphi_-)}+\cos{(\varphi_+)}\big]}{[4\alpha^2\sin^2{(\frac{\varphi_-}{2})}+1]^{1/2}},\\
=&
\frac{d^2}{8\pi^3\varepsilon_0\varepsilon R^2D}
\int {\rm d}\varphi_-\int {\rm d}\varphi_+
\frac{\cos{(\varphi_-)}+\cos{(\varphi_+)}}{[4\alpha^2\sin^2{(\frac{\varphi_-}{2})}+1]^{1/2}}.
\end{aligned}
\label{eq:potential-simplifiedfunction-2}
\end{equation}
The limits of the integration must also be changed, such that, 
%
%
%
$V_F=\frac{d^2}{8\pi^3\varepsilon_0\varepsilon R^2D}
\big[
\int_0^{2\pi}{\rm d}\varphi_-\int_{\varphi_-}^{4\pi-\varphi_-}{\rm d}\varphi_+
\frac{\cos{(\varphi_-)}+\cos{(\varphi_+)}}{[4\alpha^2\sin^2{(\frac{\varphi_-}{2})}+1]^{1/2}}
+
\int_{-2\pi}^{0}{\rm d}\varphi_-\int_{-\varphi_-}^{\varphi_-+4\pi}{\rm d}\varphi_+
\frac{\cos{(\varphi_-)}+\cos{(\varphi_+)}}{[4\alpha^2\sin^2{(\frac{\varphi_-}{2})}+1]^{1/2}}
\big]$,
and our final solution for the effective potential  becomes,
\begin{equation}
\begin{aligned}
V_F=\frac{d^2}{2\pi^3\varepsilon_0\varepsilon R^2D}\int_0^{2\pi}{\rm d}\varphi_-
\frac{\cos{(\varphi_-)}(2\pi-\varphi_-)}{[4\alpha^2\sin^2{(\frac{\varphi_-}{2})}+1]^{1/2}}.
\end{aligned}
\label{eq:potential-simplifiedfunction-newlimits}
\end{equation}

We can (and will) numerically perform the integral presented by Eq.~\ref{eq:potential-simplifiedfunction-newlimits}, but first let us examine the integral in the small ($\lim_{\alpha\rightarrow 0}$ or $R\ll D$) and large ($\lim_{\alpha\rightarrow \infty}$ or $R\gg D$) limits of $\alpha$. 
The Taylor series expansion about $\alpha =0$ for the integral in Eq.~\ref{eq:potential-simplifiedfunction-newlimits} is given to first order by:
%
$\frac{\cos{(\varphi_-)}(2\pi-\varphi_-)}{[4\alpha^2\sin^2{(\frac{\varphi_-}{2})}+1]^{1/2}}
\simeq \cos{(\varphi_-)}(2\pi-\varphi_-)\bigg(1-2\alpha^2\sin^2{\bigg(\frac{\varphi_-}{2}\bigg)}\bigg).$
Subsequently, 
 $\int_0^{2\pi}{\rm d}\varphi_-
\frac{\cos{(\varphi_-)}(2\pi-\varphi_-)}{[4\alpha^2\sin^2{(\frac{\varphi_-}{2})}+1]^{1/2}}
\simeq \pi^2\alpha^2$,
%
resulting in the effective potential:
\begin{equation}
\begin{aligned}
\lim_{\alpha\rightarrow 0}(V_F)=
\frac{d^2}{2\pi\varepsilon_0\varepsilon D^3},
\end{aligned}
\label{eq:potential-simplifiedfunction-smallalpha}
\end{equation}
which is exactly our usual dipole limit (i.e. treating the dot as a point dipole) with a square dependence on the dipole moment and an inverse cubic dependence on the separation.  

If we try to take the limit of $\alpha \rightarrow \infty$, it seems that the entire function evaluates to zero, but we must be more careful since we pass through the point $\varphi_- = 0$, resulting in $\sin^2{(\varphi_-/2)} = 0$, where the product of $\alpha^2\sin^2{(\varphi_-/2)}$ becomes finite. Thus, let us take the Taylor series expansion about $\varphi_-=0$ in Eq.~\ref{eq:potential-simplifiedfunction-newlimits}:
%
$\frac{\cos{(\varphi_-)}(2\pi-\varphi_-)}{[4\alpha^2\sin^2{(\frac{\varphi_-}{2})}+1]^{1/2}}
\simeq
\frac{2\pi-\varphi_-}{[\alpha^2\varphi_-^2+1]^{1/2}},$
 thus, $\int_0^{2\pi}{\rm d}\varphi_- \frac{\cos{(\varphi_-)}(2\pi-\varphi_-)}{[4\alpha^2\sin^2{(\frac{\varphi_-}{2})}+1]^{1/2}}
\simeq
\frac{-[4\pi^2\alpha^2+1]^{1/2} + 2\pi\alpha\sinh^{-1}{(2\pi\alpha)}+1}{\alpha^2}$,
%
for the large $\alpha$ limit. The resulting effective potential is then:
\begin{equation}
\begin{aligned}
\lim_{\alpha\rightarrow \infty}(V_F)
=&\frac{d^2}{2\pi^3\varepsilon_0\varepsilon R^3}
\times 
\\
&\bigg(\frac{-[4\pi^2\alpha^2+1]^{1/2} + 2\pi\alpha\sinh^{-1}{(2\pi\alpha)}+1}{\alpha}\bigg),
\end{aligned}
\label{eq:potential-simplifiedfunction-largealpha}
\end{equation}
which, in contrast to the small $\alpha$ limit, has an inverse polynomial (approximately cubic) dependence on the radius of the disks, rather than the distance between the disks. 

\begin{figure}[ht]
    \centering
    \includegraphics[trim = 0cm 0cm 0cm 0cm, clip=true,width=1.0\columnwidth]{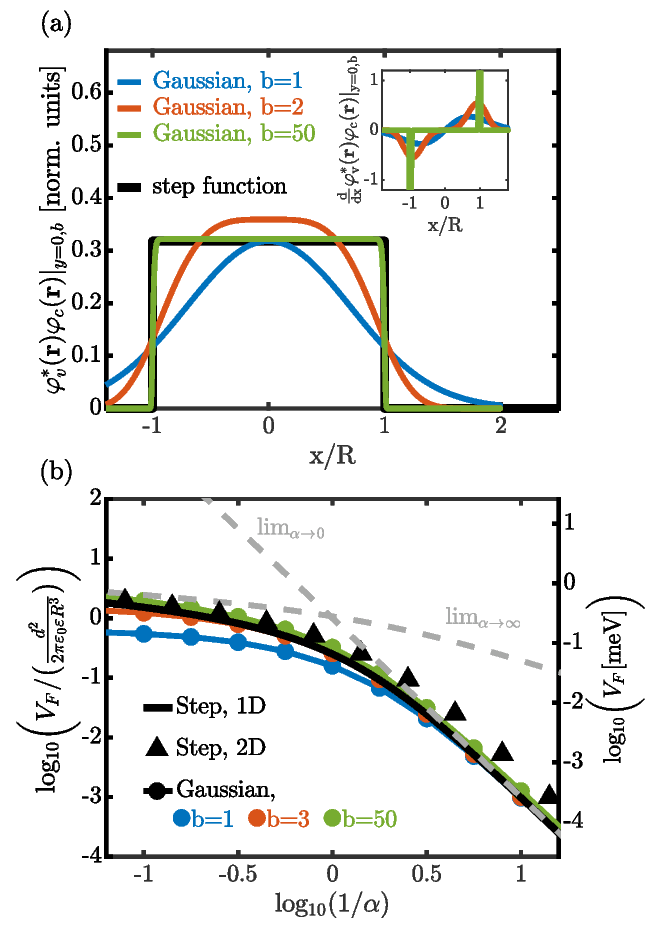} \caption{Comparison of the step function and super Gaussian wavefunctions in $x-y$ for an infinitely thin pair of quantum dots. (a) The wavefunctions and first derivative in $x$ of the wavefunctions (inset). (b) The coupling potential, $V_F$, in arbitrary units independent of radius and dipole moment (left) and in SI units for an example of $R$=10~nm, $d$=50~D, and $\varepsilon$=12. The black solid line and triangular markers refer to the step function, obtained in a 1D (Eq.~\ref{eq:potential-simplifiedfunction-newlimits}) and 2D (Eq.~\ref{eq:potential-simplifiedfunction}) integration, respectively. Grey dashed lines are the high and low $\alpha$ limits of the step function solution (Eq.~\ref{eq:potential-simplifiedfunction-largealpha} and \ref{eq:potential-simplifiedfunction-smallalpha}, respectively). Solid coloured lines with circular markers are the Gaussian solutions for $b$ = 1, 3, and 50 (blue, red, and green, respectively).  }
    \label{fig:comparehighlowalpha}
\end{figure}

Figure~\ref{fig:comparehighlowalpha} 
shows the high- and low-$\alpha$ limits (Eq.~\ref{eq:potential-simplifiedfunction-largealpha} and \ref{eq:potential-simplifiedfunction-smallalpha}, respectively) as well as the full solution to our simplified setup, provided by Eq.~\ref{eq:potential-simplifiedfunction-newlimits}. As expected, the full solution converges to the high- and low-limit solutions. 

{\it Numerical Results.---}Next, we will perform the full integration in Eq.~\ref{eq:potential-expanded2} and compare it to our simplified solution in Eq.~\ref{eq:potential-simplifiedfunction-newlimits}. 
For this, we have a few choices for simple wavefunctions: (1) step function (from previous section), (2) Gaussian (solution to the cylindrical SHO, as discussed in Stobbe \textit{et. al.} \cite{stobbe_spontaneous_2012}):

\begin{equation}
\begin{aligned}
    &(a)~~\zeta_v^*(\mathbf{r})\zeta_c(\mathbf{r}) = F(z)\frac{\Theta(R - \rho)}{\pi R^2},\\
    &(b)~~\zeta_v^*(\mathbf{r})\zeta_c(\mathbf{r}) = F(z)\frac{b}{2\pi L^2\Gamma(1/b)}e^{-(\rho^2/2L^2)^b},
\end{aligned}
\end{equation}
where $F(z)$ is either $\delta(z)$ or $\delta(z-D)$ for dot 1 or 2, respectively, $L$ is an effective length related to the disk radius by $R=\sqrt{2}L$~\cite{stobbe_spontaneous_2012}, $b$ is an integer $\geq$1, and $\Gamma(z)$ is the Gamma function for continuous variables.
The choice of $F(z)$ being a Dirac-delta function (i.e. infinitely thin dots) can be easily changed to an infinite square well solution, however, this requires as extra dimension to integrate over, so we will leave this out for now. 

The $x$ derivative of the Gaussian function is given as:
%
$\nabla_x \zeta^*_v(\mathbf{r}) \zeta_c(\mathbf{r}) 
=
\delta(z)\big(\frac{b^2}{2^b\pi L^{2(b+1)}\Gamma(\frac{1}{b})}\big)\rho^{2b-1}\cos{(\varphi)}
\exp{\big( -\frac{\rho^{2b}}{2^b L^{2b}} \big)}$,
%
which in the limit of the regular Gaussian ($b$=1), recovers the expected expression
%
$\nabla_x\zeta^*_v(\mathbf{r})\zeta_c(\mathbf{r})\vert_{b=1} 
=
\delta(z)\big(\frac{1}{2\pi L^{4}}\big)\rho\cos{(\varphi)}
\exp{\big( -\frac{\rho^{2}}{2 L^{2}} \big)}$,
%
and in the limit of $b\rightarrow\infty$, the expected expression for the derivative of a step function is also recovered:
\begin{equation}
\begin{aligned}
\lim_{b\rightarrow\infty}\nabla_x\zeta^*_v(\mathbf{r})\zeta_c(\mathbf{r}) 
=& 
\left\{
        \begin{array}{ll}
            0 & \quad \vert\rho/R\vert\neq1,  \\
            \pm\infty & \quad \rho/R = \pm 1
        \end{array}
    \right.
    .
\end{aligned}
\end{equation}

The potential can then be re-written for the Gaussian form, 
\begin{equation}
\begin{aligned}
V_F 
=& 
\frac{d^2}{4\pi\varepsilon_0\varepsilon \varepsilon}\bigg(\frac{b^2}{2^b\pi L^{2(b+1)}\Gamma(\frac{1}{b})}\bigg)^2
\int_{\rm 4D} {\rm d}\rho{\rm d}\rho^\prime{\rm d}\theta{\rm d}\theta^\prime\\
&\frac{
\rho^{2b}\rho^{\prime~2b}\cos{(\varphi)}\cos{(\varphi^\prime)}
\exp{\bigg( -\frac{\rho^{2b}}{2^b L^{2b}} \bigg)}
\exp{\bigg( -\frac{\rho^{\prime~2b}}{2^b L^{2b}} \bigg)}}
{[\rho^2+\rho^{\prime 2}-2\rho\rho^\prime\cos{(\varphi-\varphi^\prime)} + D^2]^{1/2}}.
\end{aligned}
\label{eq:matlab-int}
\end{equation}

The wavefunction envelopes and their derivative in $x$ are shown in Fig.~\ref{fig:comparehighlowalpha}(a). It is clear that the step function is recovered in the limit of $b\rightarrow\infty$. Figure~\ref{fig:comparehighlowalpha}(b) compares the solution of the step function and the Gaussian function in $x-y$ (infinitely thin in $z$), showing that the high-$b$ limit recovers the solution for the step function approximation, and in the limit of $\alpha\rightarrow 0$, all solutions converge to the dipole limit, as expected. The integrations are performed over the range of $\rho = [0,5R$] and $\theta= [0, 2\pi]$, using 200 steps (40 points per $R$), which was carefully checked for convergence.

Previously, $F(z)$ was defined as a Dirac-delta function ($\delta(z)$ and $\delta(z-D)$). To check this assumption, we also consider a finite size dot in $z$ which is defined by an infinite square well potential. Thus, the function is now $F(z)=\sqrt{\frac{\pi}{2T}}\sin{\bigg(\frac{n\pi}{T}(z+T/2)\bigg)}$ for $\vert z\vert <T/2$ and zero outside these bounds,
%
where $T$ is the thickness of the disk, and we consider only the lowest order state ($n$=1). The numerical integration is now done over two more dimensions ($z$, $z^\prime$) than before, causing simulation time to increase by a few orders of magnitude for well-converged results. For both the step function and the Gaussian solution ($b$=1), the effect of including the infinite square well potential in $z$ was investigated for a dot thickness of $R/10$, which resulted in less than a 0.5\% difference with the Dirac-delta function in $z$.

In summary, we have reported on the effective two-QD potential beyond the PDA by using a model envelope exciton wavefunction and the two-point Coulomb Green function. The model wavefunctions are calculated first for infinitesimally thin QDs with three different scenarios of in-plane envelope wavefunctions: ($i$) a circular step function, ($ii$) a Gaussian function (solution to the 2D SHO), and ($iii$) a super-Gaussian function. The resulting F\"orster coupling potential, $V_F$, was calculated in each case as a function of QD separation along the QD axis, and found that the potential agreed with the PDA limit and the infinite plate limit for very large and small separations, respectively; however, in the intermediate separation range of approximately 1/10$<$R/D$<$10, the overall potential was smaller than either limit, resulting in a screening of the usual F\"orster coupling of two point dipoles. Finally, the effect of QD thickness is included by introducing an infinite square well potential in the axial direction, and perform the full 6-D integrals, finding less than a 0.5\% deviation from the thin-QD approximations for thicknesses up to R/10. 


\section*{Funding Information}
Natural Sciences and Engineering Research Council of Canada (NSERC). 
Canadian Foundation for Innovation.
Alexander von Humboldt-Stiftung through a Humboldt Research Award.Andreas Knorr received funding from the European Unions Horizon
2020 research and innovation programme under Grant
Agreement No. 734690 (SONAR).
We thank Marek Korkusinski
and Dan Dalacu for useful discussions.

\bibliography{Refs}

\end{document}